\documentclass{webofc}
\usepackage[varg]{txfonts}   
\usepackage{amssymb,amsmath,bm,mathtools}

\begin{document}

%
%

\title{The Jefferson Lab Positron Physics Program}
%
%

\author{\lastname{Eric} \lastname{Voutier}\inst{1}\fnsep\thanks{\email{voutier@ijclab.in2p3.fr}} for
        \lastname{The Jefferson Lab Positron Working Group}\thanks{\email{pwg@jlab.org}} }

\institute{Universit\'e Paris-Saclay, CNRS/IN2P3, IJCLab, 91405 Orsay, France 
}

\abstract{The Ce$^+$BAF project at the Thomas Jefferson National Accelerator Facility intends to develop over the coming years a  high-duty cycle, high intensity, and high polarization positron beam to serve a unique Nuclear Physics experimental program. It generically comprises the study of the effects of the two- and multi-photon exchange mechanisms, the investigation of the nuclear structure at the nucleonic and partonic scales, and tests of the standard model of particle physics and interactions. This proceedings illustrates this physics program through specific examples and presents the Ce$^+$BAF positron injector concept.}
\maketitle

%
%

\section{Introduction}
\label{intro}

The perspective of high duty-cycle and high intensity polarized and unpolarized positron beams, in complement to the existing CEBAF (Continuous Electron Beam Accelerator Facility) 12~GeV electron beams, has been nurtured since the very first 6~GeV upgrade of the CEBAF accelerator. Over the years, experimental results about the Electromagnetic Form Factors (EFFs) and the Generalized Parton Distributions (GPDs) of the nucleon pointed towards the importance of positron beams in determining these fundamental quantities of the nucleon structure~\cite{Vou14}. Further ideas emerged about testing the predictions of the standard model~\cite{Zhe21,Fur21}, exploring the dark matter sector~\cite{Bat21}, or investigating electroweak processes~\cite{Mel21}. A sustained and extensive research effort has evolved both in the physics~\cite{Acc21} and technical~\cite{Gra23} domains, aiming to assess the potential of an experimental program and address the technological challenges associated with high duty cycle positron beams. The high scientific value of a positron program at Jefferson Lab (JLab) has been recognized by the Program Advisory Committee such that the development of positron beam capabilities is now identified as the first step of the future CEBAF upgrade.

The Jefferson Lab Positron Working Group (JLab PWG) recently published the JLab Positron White Paper (PWP) as a topical issue of The European Physics Journal, gathering a non-exhaustive but yet impressive list of physics motivations and experimental scenarios for positron beams at CEBAF~\cite{Ala22}. The program is organized around three essential pillars: two- and multi-photon exchange physics, nucleon and nuclei structure, and tests of the standard model. The present conference proceedings highlights some of the flagship examples from the PWP, showcasing the benefit of positron beams for expanding the CEBAF physics reach. Additionally, it also discusses the current Ce$^+$BAF concept for producing and implementing positron beams at CEBAF.

%
%

\section{Electromagnetic form factors}
\label{sec-2}

Following the first measurements of the polarization transfer of a longitudinally polarized electron beam to the recoil protons in the elastic electron scattering process~\cite{Jon00}, the validity of the One-Photon Exchange (OPE), Born approximation, for the description of the electromagnetic interaction at high four-momentun transfer $Q^2$ was questioned. Indeed, it was suggested that mechanisms associated with Two-(hard) Photon Exchanges (TPE) may reconciliate the measurements of the ratio of the proton EFFs as obtained from Rosenbluth separation or polarization transfer  experiments~\cite{Gui03,Blu03}. The TPE hypothesis complicates the description of the electromagnetic structure of the nucleon which would then involve 3 generalized form factors corresponding to 8 unknown $Q^2$-dependent quantities to be determined from experimental data. Therefore, unpolarized and polarized electron scattering observables solely can no longer resolve this structure. This feature is even more embarassing because of the model dependence of TPE calculations. The experimental assessment of these effects is consequently mandatory for the understanding of the nucleon structure. In that respect, unpolarized and polarized positron beams at CEBAF have the unique opportunity to bring a definitive answer about the TPE hypothesis, because of the sensitivity of the elastic cross section to the lepton beam charge.

\begin{figure}[!h]
\centering
\includegraphics[width=0.90\textwidth,clip]{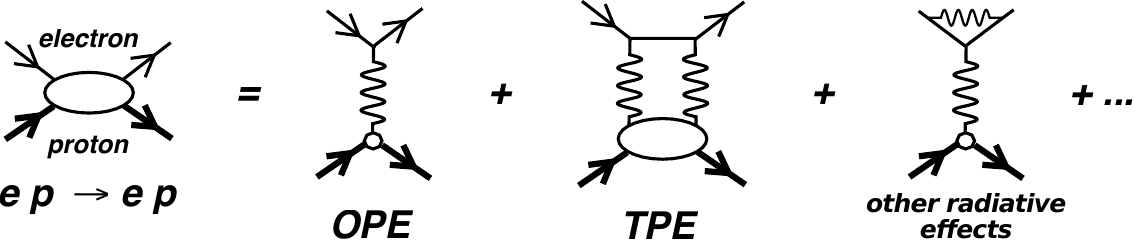}
\caption{Feynmann diagram series for elastic $ep$ scattering featuring: the One-Photon Exchange (OPE) mechanism, the Two-Photon Exchange (TPE) one, and other radiative effects.}
\label{fig-1}  
\end{figure}

The elastic $e^{\pm}p$ cross section can be built from the coherent sum of several elementary reaction amplitudes (Fig.~\ref{fig-1}). Particularly, the quantum interference between the OPE and TPE mechanisms changes sign with the lepton beam charge ($\pm e$). This is expressed in the elastic reduced cross section $\sigma_R$ characterizing the non-point like structure of the proton~\cite{Gui03}
\begin{equation}
\sigma_R = G_M^2 + \frac{\epsilon}{\tau} G_E^2 - 2 e \, G_M \Re \mathrm{e} \left[ \mathcal{F}_0 \right] - 2 e \, \frac{\epsilon}{\tau} G_E \Re \mathrm{e} \left[ \mathcal{F}_1 \right] + \mathcal{O}(\alpha^4) \equiv \sigma_{1\gamma} - e \, \sigma_{1\gamma:2\gamma} + \sigma_{2\gamma} 
\end{equation}
where $\sigma_{1\gamma}$ is the Born contribution to the reduced cross section, $\sigma_{1\gamma:2\gamma}$ is the OPE-TPE interference contribution, and $\sigma_{2\gamma}$ is the negligible pure TPE contribution; $G_E$ and $G_M$ are the electric and magnetic form factors of the proton, $\epsilon$=${\left[ 1 + 2(1+\tau)\tan^2(\theta/2)\right]}^{-1}$ is the polarization of the virtual photon with $\theta$ the angle of the scattered electron and $\tau$=$Q^2/4M^2$ with $M$ the proton mass; $\left(\Re \mathrm{e} \left[ \mathcal{F}_0 \right],\Re \mathrm{e} \left[ \mathcal{F}_1 \right] \right)$ parameterize the TPE contribution to $\sigma_R$. Comparing elastic cross sections measured with lepton beams of opposite charges, it is thus possible to isolate and measure the TPE contribution from $\sigma_{1\gamma:2\gamma}$. \newline
It is the purpose of the PR12+23-008 experiment~\cite{Ber21,Sch23} to measure the ratio $R_{2\gamma}$ of the positron to electron cross sections exploring a wide range of the $(\epsilon,Q^2)$ phase space, particularly accessing regions where TPE are predicted to be significant. This is an invaluable benefit of Ce$^+$BAF as compared to previous attempts to measure TPE~\cite{Adi15,Rac15,Hen17} which lacked the necessary kinematical reach to draw meaningful conclusions. The PR12+23-012 experiment~\cite{Arr21,Nyc23} proposes an alternative approach involving the comparison of the $\epsilon$-slope of $\sigma_R$ as obtained from a Super-Rosenbluth separation usind $e^{\pm}$ beams. This complementary technique features a larger sensitivity to TPE and provides insights into other eventual effects which may contribute to the form factor ratio disagreement. Conversely to cross sections, polarization observables are predicted to be much less affected by TPE~\cite{Gui03}. The LOI12+23-008 experiment~\cite{Puc21,Puc23} aims at testing this prediction by measuring the polarization of recoil protons in the elastic scattering of longitudinally polarized electrons and positrons off hydrogen. This set of measurements will definitely establish the evidence or absence of TPE in elastic scattering.

However, TPE and more generally multi-photon effects are not specific of the nucleon nor of elastic scattering. Considering the non-zero single spin asymmetries measured by several collaborations on different nuclei, it is legitimate to question the  significance of TPE in the elastic scattering off nuclei~\cite{Kut22}, as well as in the Deep Inelastic Scattering (DIS) regime. Inclusive scattering off nuclei is also known to suffer from multi-(soft) photon exchanges which result in a distorsion of the incoming and scattered electron wave-function under the influence of the $Z$-Coulomb field of a nucleus. The precise measurement of these effects is of importance for the determination of nuclear radii and potentially the Coulomb sum rule~\cite{Gue23}. The Coulomb field of heavy nuclei is also known to be responsible of the acceleration of the incoming and outgoing electrons. These effects are usually taken into account using the prescriptions of the improved Effective Momentum Approximation established in the quasi-elastic regime~\cite{Ast08}. There is no clear guidance for DIS except that these effects have been shown sizeable and of particular importance for the understanding of the EMC (European Muon Collaboration) effect~\cite{Gas23}. Thanks to the beam intensity, high duty cycle and large spin polarization, the Ce$^+$BAF project can uniquely address these issues and bring the necessary inputs required for precision physics.

%
%

\section{Generalized parton distributions}
\label{sec-3}

The GPDs paradigm~\cite{Mul94} has profoundly renewed the understanding of the nucleon structure. As describing the correlations between partons, GPDs allow us to access static and dynamical information about the nucleon structure, ultimately learning about the mechanics of Quantum Chromodynamics. This comprises the total angular momentum of the nucleon carried by the partons~\cite{Ji97}, the distribution of forces experienced by partons inside the nucleon~\cite{Pol03,Bur20} or the Gravitational Form Factors (GFFs) of the nucleon~\cite{Ji97,Ter16}. These dynamical properties of the nucleon are obtained through integrals of GPDs, calling therefore for a comprehensive experimental program exploring the full physics phase-space and providing sensitive enough observables to determine unambiguously the different nucleon GPDs.

\begin{figure}[!t]
\begin{center}
\includegraphics[width=0.85\textwidth]{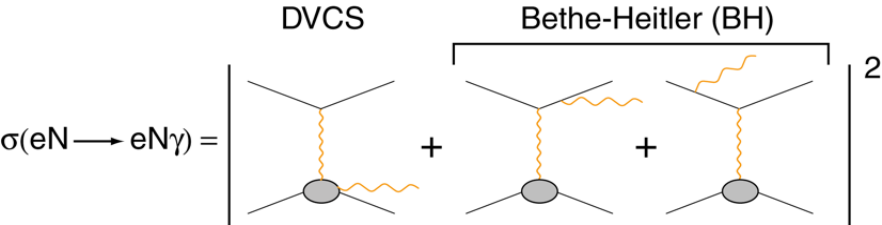}
\caption{Lowest QED-order amplitude of the $eN\gamma$ reaction.}
\label{EAgamma}
\end{center}
\end{figure}

Deeply Virtual Compton Scattering (DVCS), corresponding to the production of a real photon from a parton of the nucleon or the nucleus ($e^{\pm}N\gamma$), is the golden chanel to access GPDs. The $e^{\pm}N\gamma$ reaction (Fig.~\ref{EAgamma}) gets also contributions from the Bethe-Heitler (BH) process where the real photon is produced either by the incoming or the outgoing electron, such that the $e^{\pm}N\gamma$ cross section is the coherent sum of the reaction amplitudes
\begin{equation}
d^5 \sigma_{\lambda,0}^e = d^5 \sigma_{\mathrm{BH}} + d^5 \sigma_{\mathrm{DVCS}} + \lambda \, d^5 \widetilde{\sigma}_{\mathrm{DVCS}} - e \left[ d^5 \sigma_{\mathrm{INT}} + \lambda \, d^5 \widetilde{\sigma}_{\mathrm{INT}} \right] \label{eNg}
\end{equation}
where INT denotes the contribution of the interference amplitude between the DVCS and BH processes. The previous Eq.~\eqref{eNg} relation singles out the lepton beam polarization ($\lambda$) and charge dependence of the cross section. Similarly to elastic scattering, the comparison between electron and positron cross sections separate the interference contributions which can be further decomposed taking advantage of the beam polarization sensitivity. Thus the comparison between polarized lepton beams of opposite charges allows for a full separation of the 4 unknwon amplitudes of the $e^{\pm}N\gamma$ cross section involving GPDs, the pure BH amplitude being known from the electromagnetic form factors. The same feature occurs with a polarized target where the $e^{\pm} \vec{N} \gamma$ cross section can be written 
\begin{equation}
d^5 \sigma_{\lambda,P}^e =d^5 \sigma_{\lambda,0}^e + P \, d^5 \Delta\sigma_{\lambda}^e
\end{equation}
where $P$ represents the target polarization and with 
\begin{equation}
d^5 \Delta \sigma_{\lambda,0}^e = \lambda \, d^5 \Delta\sigma_{\mathrm{BH}} + \lambda \, d^5 \Delta\sigma_{\mathrm{DVCS}} + d^5 \Delta\widetilde{\sigma}_{\mathrm{DVCS}} - e \left[ \lambda \, d^5 \Delta\sigma_{\mathrm{INT}} + d^5 \Delta\widetilde{\sigma}_{\mathrm{INT}} \right] \, .
\end{equation}
The experimental knowledge of these different amplitudes is of particular importance since they access different nature and combination of GPDs: the DVCS amplitudes are proportional to a bilinear combination of Compton Form Factors (CFFs) which real parts $(d^5 \sigma_{\mathrm{DVCS}},d^5 \Delta\sigma_{\mathrm{DVCS}})$ involve the Cauchy principal value of GPDs integral, while the imaginary parts $(d^5 \widetilde{\sigma}_{\mathrm{DVCS}},d^5 \Delta\widetilde{\sigma}_{\mathrm{DVCS}})$ feature the GPDs value at a fixed phase-space point; similarly, the INT amplitudes are proportional to a linear CFFs combination; both the DVCS and INT amplitudes access different CFFs combination depending on the target polarization. The resolution of the partonic  structure of hadrons requires such a diverse experimental observable landscape. \newline
Along these lines, the PR12+23-006 experiment~\cite{Afa21,Mun23} aims at very precise measurements of the $e^{+}p\gamma$ unpolarized cross section at selected kinematics where $e^{-}p\gamma$ data are currently being acquired using the Neutral Particle Spectrometer (NPS)~\cite{Mun13}. The PR12+23-002 experiment~\cite{Bur21,Vou23} intends to use polarized electron and positron beams to determine Beam Charge Asymmetry (BCA) observables on the proton with the CLAS12 spectrometer~\cite{Bur20}: the unpolarized BCA sensitive to the real part of the CFFs; the polarized BCA measuring the imaginary part of the CFFs differently from a Beam Spin Asymmetry (BSA) but carrying the same physics content; the charge and polarization averaged BCA signing the presence of eventual higher twist effects. These measurements have been shown to strongly impact the determination of the real part of the CFF $\mathcal{H}$, improving the accuracy of its determination by reducing the correlations bewteen the nucleon CFFs built in observables~\cite{Dut21}. The extension of DVCS measurements on a neutron have also been investigated~\cite{Nic21} in the perspective of a quark-flavor separation of the GPDs and considering its unique access to the $E$ GPD, this new piece of information about the nucleon structure, of interest for the nucleon spin sum rule and inaccessible to DIS. Further reaction chanels have been proposed for the study of the partonic structure of the nucleus as measured by DVCS~\cite{Fuc21} and the exploration of the physics phase-space of GPDs through Double Deeeply Virtual Compton Scattering (DDVCS)~\cite{Zha21}. This represents a comprehensive experimental program providing unique measurements of the real part of CFFs and opening a new line of research about the GFFs of nucleons and nuclei.

%
%

\section{Tests of the standard model}
\label{sec-4}

While the two previous class of measurements illustrate the interest of positron beams in reactions involving more than one  Quantum Electro-Dynamics (QED) based amplitude, another class of experiments takes advantage of the peculiarities of positrons to investigate new mechanisms or access the neutral- and charged-currents of the electroweak sector. In these respects, positron beams provide new and original ways to investigate the existence of Physics Beyond the Standard Model (PBSM). 

\begin{figure}[!h]
\begin{center}
\includegraphics[width=0.320\textwidth]{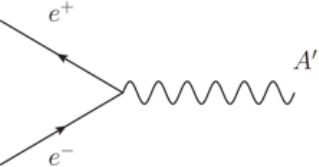}
\caption{Annihilation in-flight of a positron into an $A'$-dark photon.}
\label{PosLDM}
\end{center}
\end{figure}
Considering the search for Light Dark Matter (LDM) through the production of a feebly interacting boson, the so-called $A'$-dark photon, positron beams procure a new chanel of exploration different from the $A'$-strahlung investigated at several facilities.  The annihilation in-flight of positrons (Fig.~\ref{PosLDM}) allows for the direct production of the $A'$ which may be thought more sensitive to LDM than the $A'$-strahlung if responsible of the anomalies observed in cosmic radiations. An experimental  setup measuring the energy deposit at each interaction would sign the $A'$ production with a deposit smaller than the beam 
energy. The experimental projections of this chanel at Ce$^+$BAF are remarkable with a significant exclusion power with respect to theoretical dark matter scenarios~\cite{Bat21}. \newline
Another chanel of investigation of PBSM is the lepton flavor violation in the charged lepton sector, which observation has eluded all experimental searches to date. The reaction $e^+ N \to \mu^+ X$ in the DIS regime is capable to test the existence of  leptoquarks candidates different from those accessed with an electron beam. The Ce$^+$BAF positron beam together with the SoLID~\cite{Arr23} spectrometer upgraded for muon detection is predicted to significantly improved the mass limits measured at HERA~\cite{Fur21}. An even more challenging experimental measurement concerns the access to the axial-axial electroweak neutral current coupling $C_{3q}$ which was never measured with $e^{\pm}$ beams~\cite{Zhe21-1}. Such a measurement which would provide a test of standard model predictions, is considered as a second generation measurement because extremely demanding in terms of beam properties, detector control, and understanding of QED effects.

%
%

\section{The Ce$^+$BAF positron beam concept}
\label{sec-5}

\begin{figure}[!t]
\begin{center}
\includegraphics[width=0.80\textwidth]{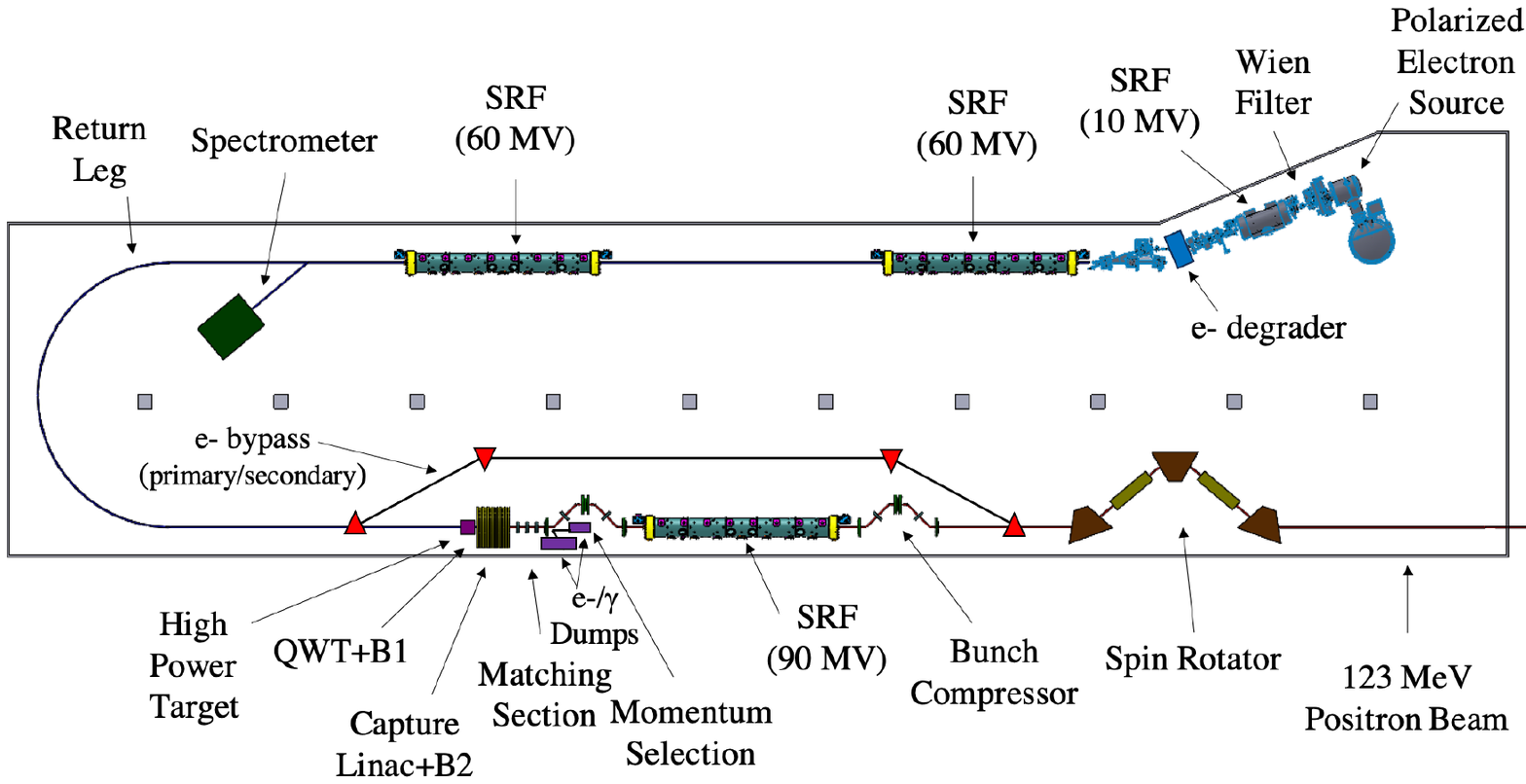}
\vspace*{-20pt}
\caption{The Ce$^+$BAF positron injector concept at LERF together with the primary/secondary electron by-pass line~\cite{Gra23}.}
\label{PosInj}
\end{center}
\end{figure}

The novelty of the Ce$^+$BAF positron source and injector consists essentially in the high duty cycle, high intensity, and high polarization of the positron beam for a linear accelerator. The basic principle of the positron source is based on the PEPPo (Polarized Electrons for Polarized Positrons) principle which was demonstrated at the CEBAF injector to produce up to 82\% positron polarization~\cite{Abb16}. The polarized positrons at PEPPo are obtained from polarized electrons interacting within a tungsten target. The longitudinal polarization of the electrons transfers into circular polarization of the bremsstrahlung photons which then transfers into longitudinal and  transverse polarization components of the leptons produced within the same target by the $e^+e^-$-pair creation process. The implementation of PEPPo into the Ce$^+$BAF injector~\cite{Gra23} asks for a high intensity polarized electron source ($\sim$1~mA), a high enough electron beam energy (100-150~MeV) to access efficient production rates, and the selection of the positron momentum to operate either with high or low polarization, correspondingly moderate or high positron beam intensity~\cite{Hab23,Hab24}. The positron source develops from this initial electron beam and successively involves~\cite{Hab22}: a high power target capable of absorbing a 17~kW power, a collection system composed of a high-field solenoid and a warm RF (Radio Frequence) section embedded inside a low field solenoid, a momentum selection chicane, a cold RF section bringing the selected positrons at 123~MeV/$c$, and a bunch length compression section. This line continues with a spin rotator which prepares the beam for maximum polarization in the experimental hall. The full concept shown on Fig.~\ref{PosInj} also comprises a by-pass line designed for electron beam delivery. It  concerns either primary electrons produced at the polarized electron source or secondary electrons produced at the electron degrader at the end of the 10~MeV electron injector. The degrader aims at creating electron beams with emittance and momentum spread comparable to those of the positron beam, as required by experiments asking  alternate operation positron and electron beams.

The Ce$^+$BAF injector would be installed in the LERF (Low Energy Recirculator Facility) building which was previously hosting the JLab's Free-Electron Laser (FEL) program. The existence of such a building already equipped with electric and cryogenic capabilities and safety systems is an invaluable benefit for the development and installation of Ce$^+$BAF. A strong R\&D program is currently taking place, particularly addressing the two high-risk components of the positron source that are the high intensity polarized electron source and the high power target~\cite{Ush23}.

%
%

\section{Summary}
\label{summa}

The perspective of positron beams at CEBAF opens new avenues to investigate the nuclear structure at the nucleonic and partonic scales. It also offers novel ways to search for new physics or test the predictions of the standard model. The physics scenarios described in these proceedings are selected examples and we refer the reader to the PWP~\cite{Ala22} for a more complete and elaborated but still not exhaustive list.

%
%

\section*{Acknowledgments}
\label{ack}

This work has received funding from the European Union's Horizon 2020 research and innovation program under grant agreement No 824093 and was supported in part by the U.S. Department of Energy, Office of Science, Office of Nuclear Physics under contract  DE-AC05-06OR23177, and the French Centre National de la Recherche Scientifique.

\end{document}